\input phyzzx
\hoffset=0.375in
\overfullrule=0pt

\def\au{{\rm AU}}

\def\max{{\rm max}}
\def\min{{\rm min}}
\def\lim{{\rm lim}}

\def\nep{{\rm nep}}

\def\pc{{\rm pc}}

\def\kms{{\rm km}\,{\rm s}^{-1}}
\twelvepoint
\font\bigfont=cmr17
\centerline{\bigfont No Death Star -- For Now}
\bigskip
\centerline{{\bf Jay A.\ Frogel} and 
{\bf Andrew Gould}\footnote{1}{Alfred P.\ Sloan Foundation Fellow}
}
\smallskip
\centerline{Department of Astronomy, Ohio State University, Columbus, OH 43210}
\smallskip
\centerline{frogel@astronomy.ohio-state.edu, gould@astronomy.ohio-state.edu}
\bigskip
\centerline{\bf Abstract}
\singlespace 

	A star passing within $\sim 10^4\,\au$ of the Sun would 
trigger a comet shower that would reach the inner solar system about 0.18 Myr
later.  We calculate a prior probability of $\sim 0.4\%$ that a star 
has passed  this close to the
Sun but that the comet shower has not yet reached the Earth.
We search the HIPPARCOS catalog for such recent 
close-encounter candidates and, in agreement with Garc\'\i a-S\'anchez et al.\
(1997), find none.  The new result reported in this {\it Letter} is an 
estimation of the completeness of the search.  Because of the relatively bright
completeness limit of the catalog itself, $V\sim 8$, the search is
sensitive to only about half the stars that could have had such a near 
encounter. On the other hand, we show that the search is sensitive to nearly
all of the {\it past} encounters that would lead to a {\it major} shower in 
the  {\it future} and conclude that it is highly unlikely that one will occur 
during the next 0.5 Myr.

\bigskip
Subject Headings: astrometry -- comets: general -- stars: kinematics

\endpage
\chapter{Introduction}

	Although most new long-period comets that enter the inner solar
system have semi-major axes
$a\gsim 2\times 10^4\,\au$, simulations show that the same scattering processes
responsible for creating the reservoir of these comets (the ``Oort Cloud'') 
also generated a region with a much higher density of comets at $3\times
10^3\,\au\lsim a\lsim 2\times 10^4\,\au$, 
(Hills 1981; Duncan, Quinn, \& Tremaine 1987).  
When a star has an impact parameter $b\sim 10^4\,\au$, and so passes 
directly through this
much more densely populated inner Oort Cloud, 
the number of new comets entering the inner solar system can then increase by 
up to
40-fold for $\sim 3$ Myr until interactions with planets again clear
the loss cone (Duncan et al.\ 1987, but also see Weissman 1996a,b, 1997).  
Such a comet shower might result in a
large increase in impact frequency on Earth; these multiple impacts 
over a relatively short time interval could
cause stepwise mass extinctions of life
(Hut et al.\ 1987).
Such violent encounters occur rarely,  at average intervals
$$t_i \sim (\pi b^2 n \bar v)^{-1}\sim 45\,{\rm Myr}\eqn\interv$$
where $n\sim 7.4\times 10^{-2}\,\pc^{-3}$ is the local density of stars
(Gould, Bahcall, \& Flynn 1996, 1997; Wielen, Jahreiss, \& Kr\"uger 1983; 
Liebert, Dahn, \& Monet 1988), 
and $\bar v\sim 38\,\kms$ is the mean speed of the Sun relative
to passing stars.  Thus, of order half of all new comets (and hence half
of catastrophic cometary impacts on Earth) occur in a steady drizzle while the
other half occur during brief 3 Myr episodes triggered by the rare passage
of a ``death star''.

	The present epoch is one of steady drizzle (Weissman 1993, 1996b).  
It is possible, however,
that a death star has already passed close to the Sun and that a comet shower 
will commence in the near future.
When a star passes the Sun at
$b\sim 10^4\,\au$, the perturbed comets (with $a\sim b/2$) do not
reach the Earth until after a delay time
$$t_d = {1\over 2}\biggl({a\over \au}\biggr)^{3/2}{\rm yr}\sim 0.18\,{\rm Myr}.
\eqn\tdelay$$
Hence, there is a small, but non-negligible chance, $t_d/t_i\sim 0.4\%$ that
a death star has already passed the Sun but its rain of death 
and destruction has not yet begun.

	One could easily test for this possibility if accurate distances, $d$,
proper motions $\mu$, and radial velocities, $v_r$, were available for all
nearby stars.  For close encounters, $b\ll d$, the impact parameter is
given by
$$ b = {\mu d^2\over v_r}.\eqn\bmudv$$
Unfortunately, no such complete inventory is available.  However, the
HIPPARCOS satellite does provide accurate distances and proper motions
for a stellar sample that is nearly complete to $V=8$.  In \S\ 2 we 
search for death star candidates in the HIPPARCOS sample.  
Garc\'\i a-S\'anchez et al.\
(1997) have conducted a similar search and found one star that may
encounter the Sun about 1 Myr in the future.  
M\"ull\"ari 
\& Orlov (1996) have also conducted such searches with pre-HIPPARCOS
data and identified some of the same stars as candidates for close encounters.
What is new about the
present study is that we evaluate the completeness of our search (\S\ 3)
for stars that recently passed close to the Sun (or are about to do so)
and, more importantly, the completeness
for recent encounters that could lead to a major comet shower in the 
``near'' future.  In \S\ 5, we rule out such a major shower at the 96\% 
confidence level.

\chapter{Death Star Search}

	The HIPPARCOS catalog (ESA 1997) is virtually complete for $V<7.3$
and is $\sim 70\%$ complete at $V=8$. For $V>8$, the completeness drops 
rapidly, particularly for death star candidates.
Among the numerous stars at $V\sim 10$, only the nearby ones
could have recently passed the Sun.
The primary method for identifying
nearby stars is from their high proper motion, whereas a death star is 
characterized precisely by its {\it low} proper motion and so would escape
this selection technique and would most likely not be included in the HIPPARCOS
catalogue.  
We therefore restrict our primary search to stars $V\leq V_\max = 8$.  
Next, we restrict attention to close encounters, with impact
parameters $b<b_\max= 0.05\,\pc$ ($10^4\,\au)$, thus ensuring passage through 
the inner Oort Cloud.  

	Equation \bmudv\ can be then be rewritten $\mu<b_\max v_r/d^2$.
We assume that $v_r<v_\max = 150 \,\kms$, based on the
following considerations.  First, it includes all disk stars and
virtually all thick disk stars (see \S\ 3).  
While this will exclude most spheroid stars,
these comprise only $\sim 0.2\%$ of main-sequence stars in the solar 
neighborhood (Gould, Flynn, \& Bahcall 1998).  Even taking account of their
higher speed, they comprise only $\sim 1\%$ of the flux of nearby stars.
Second, with this limit, we need to search for stars only within a distance
$$d < d_\max = v_\max t_d = 27\,\pc\eqn\dlimit$$
from the Sun.  At this limiting distance, it is necessary to evaluate whether 
the star's proper motion satisfies $\mu < b_\max/(v_\max t_d^2) = 2\,\rm mas\,
yr^{-1}$.  This is already close to the size of the HIPPARCOS errors. 
A doubling of $v_\max$, would necessitate a search to twice the
distance where the expected proper motions would be only half as big.  Hence,
any detected candidates would be of marginal significance.

	Thus, our search is restricted by
$$V_\max = 8,\quad d_\max = 27\,\pc,\quad v_\max = 150\,\kms,\quad
b_\max = 0.05\,\pc.\eqn\parmlims$$
For the 753 HIPPARCOS stars that satisfy the first two conditions in 
equation \parmlims, parallax errors are generally only a few percent.
However, proper motion errors can be a large fraction of the allowable
proper motion, 
$$\mu< \mu_\max (d) = {b_\max v_\max\over d^2} = 15.8\,
\biggl({d\over 10\,\pc}\biggr)^{-2}\,\rm mas\,yr^{-1}.\eqn\mumax$$
In order to allow for proper-motion errors, we adopt a lower-limit estimate
$\mu_{\rm low}^2 = 
(\max\{0,|\mu_\alpha|-1.5\sigma_\alpha\})^2 +
(\max\{0,|\mu_\delta|-1.5\sigma_\delta\})^2,$
where $\mu_\alpha$ and $\mu_\delta$ are the two components of the proper
motion, and $\sigma_\alpha$ and $\sigma_\delta$ are their errors.
This search yields two candidates, HIPPARCOS 
24502 and 35550.  However, both of these
stars are among the handful (17 out of the 753 or 2\%) with proper-motion 
errors exceeding 5 mas $\rm yr^{-1}$ 
in at least one direction.  Indeed their errors (in mas $\rm yr^{-1}$)
are (23.72, 12.19) and (11.31, 5.00) respectively. Elimination of these 17
stars leaves no candidates that satisfy our search criteria.

	HIPPARCOS proper motions are relative to a non-rotating frame. 	
Hence, there are no Coriolis force corrections to our analysis.
However, 
Galactic tides will cause the impact parameter to deviate from
equation \bmudv\ by \hfil\break
$|\Delta{\bf b}| \simeq(\pi/3)G\rho t^2 d |\sin (2b_g)|$ where $\rho$ is
the local disk density, $t=d/v_r$, 
and $b_g$ is the Galactic latitude of the passing
star.  For relevant parameters ($\rho\sim 0.1\,M_\odot\,\pc^{-3}$,
$t<$0.5 Myr, $d<27\,\pc$), $|\Delta{\bf b}|<700\,\au$ which can be ignored.

\chapter{Completeness of the Basic Search}

	Suppose that a death star had passed the Sun during the past 
$t_d = 0.18\,$Myr.  What is the probability that we would have detected it
in our search of the HIPPARCOS catalog? We first
first construct a luminosity function of dwarf stars by combining the
data from Wielen et al.\ (1983) for $4\leq M_V \leq 10$ and from Gould
et al.\ (1997) for $11\leq M_V\leq 18$.  Note that the latter is uncorrected
for binaries which is what is desired in the present context because the
overwhelming majority of binaries are close enough that they would act like
single-star perturbers.  We obtain a luminosity function 
in units of $10^{-3}\,\rm pc^{-3}\,mag^{-1}$ over the range $4\leq M_V\leq 18$ 
of $\Phi = $ 
(2.3, 3.3, 4.0, 3.0, 4.0, 4.6, 7.1, 9.5, 10.1, 6.3, 3.9, 1.9,
1.7, 2.3, 2.3).  
A rigorous treatment of brighter stars, $M_V<4$, would be complicated;
but since their relative number is small, we use an approximate treatment. 
Main-sequence stars brighter than $M_V=-0.6$ can be ignored since there 
are none within $d_\max=27\ 
\pc$ of the Sun.  For the remaining early-type stars, we
adopt a total number density of
$3\times 10^{-3}\,\pc^{-3}$, and assume that their velocity distribution
is like that of other disk stars, but without a thick-disk component.
The total main-sequence
number density is then $n_{ms}\sim 69\times 10^{-3}\,\pc^{-3}$.
The number density of giants ($M_V<2.2$) is $n_g=2\times 10^{-3}\,\pc^{-3}$ 
(C.\ Flynn 1998, private communication) and
that of white dwarfs is $n_{wd}\sim 3\times 10^{-3}\,\pc^{-3}$
(Liebert et al.\ 1988).  Hence, the total number
density is $n = 74\times 10^{-3}\,\pc^{-3}$.

	To evaluate the completeness we perform the following Monte Carlo
simulation.  First, we model the local stellar
population as composed 90\% of disk stars with mean velocity relative to the
Sun in Galactic coordinates of (9, 20, 7)$\,\kms$ and with dispersions 
(38, 19, 20)$\,\kms$ and 10\% of thick disk stars with 
mean velocity (9, 62, 7)$\,\kms$ and with dispersions (66, 37, 38)$\,\kms$ 
(Casertano, Ratnatunga, \& Bahcall 1990).  We begin the Monte Carlo by drawing
a stellar velocity randomly from this distribution.  Next, we choose a random
position on the sky and calculate $v_r$.  Then we choose $b^2$ randomly
from the interval $[0,b_\max^2]$.  If $b v_r > b_\max v_\max$ then the
star would not be included in our sample even though $b<b_\max$.  
Otherwise we perform the following calculation for
each apparent magnitude.  First we find $d_\lim$,
the limiting distance to which the
star could be detected.  This is the lesser of $d_\max=27\,\pc$ (the
distance limit of the study) and the distance at which the star's apparent
magnitude would rise above $V_\max = 8$ (the magnitude limit).  At this
limiting distance, the elapsed time since closest approach to the Sun is 
$t_\lim=d_\lim/v_r$.
If $t_d<t_\lim$, then the star is always detected.
Otherwise it is detected only if the passage is sufficiently recent, that is
a fraction $t_\lim/t_d$ of the time.  Each simulated event is weighted
by $|v_r|$ since the flux of encounters is proportional to the relative
velocity.  

	Essentially all encounters of stars $M_V<6$ are detected, so the
details of the luminosity functions for giants and early main sequence stars
are not important; only their total number density.  
White dwarfs are all assigned a magnitude
$M_V=15$ and, of course, very few ($\sim 6\%$) are detected.

\FIG\one{
Percentage of all close encounters 
in the Monte Carlo simulation as a function of 
stellar type.  The three bars to the left represent white dwarfs, giants,
and early main-sequence stars ($M_V<4$).  The 15 bars to the right represent
one
magnitude bins of the main-sequence luminosity function.  The height of each
bar is the percentage of all close encounters
due to that type of star.  The black
portion of the bar represents the fraction of these events that would have been
detected in our search of the HIPPARCOS catalog.
}

	Because the underlying catalog is complete to $V=7.3$ and only
70\% complete at $V=8$, we repeat the entire calculation with $V_\max=7.3$
and take as our final result the average of the two calculations weighted
70:30. Figure \one\ shows the completeness as a function of absolute magnitude.
The completeness for the stellar population as a whole is 53\%.  Taking
account of the fact that we are insensitive to most spheroid stars and that we
eliminated 2\% of the stars because of the poor precision of their proper
motions, this estimate is reduced to 51\%.

	Of course there is no hard boundary at $b=0.05\,\pc$ separating
catastrophic encounters from more benign ones.  One might then ask how 
sensitive we are to encounters at somewhat larger impact parameters.  To
answer this question, we increase $b_\max$ by a factor 2 to 0.1 pc.  
Correspondingly, we increase $t_d$ by $2^{3/2}$ to 0.5 Myr.  We then repeat
the same completeness calculation and find 34\%.  After again taking account
of insensitivity to spheroid stars and the poor-precision stars, this is
reduced to 33\%.

\chapter{A Candidate}

	In order to evaluate the completeness of our study, we imposed
a magnitude limit $V<8$.  Nevertheless, it is of interest to ask whether
any of the stars in the HIPPARCOS catalog with $V>8$ are death star candidates.
We therefore ran our search on the entire catalog (but excluding stars with
proper-motion errors exceeding $5\,\rm mas\,yr^{-2}$) and
found only one candidate:  HIPPARCOS 89825 (Gl 710)
the same star found by Garc\'\i a-S\'anchez  et al.\ (1997).
Since Gl 710's $v_r = -14.3\,\kms$ (Stauffer \& Hartman 1986), its rendezvous
with the Sun is in the future, not the past.  
The impact parameter is $b = 0.33\pm 0.15\,\pc$.
Thus, in about 1.3 Myr, this star will have a close encounter with the Sun.
Weissman et al.\ (1997) estimate that this will cause a 
(probably not significant)
increase of 50\% in the long-period comet flux.

\chapter{Sensitivity to High-Damage Encounters}

	As we discussed in \S\ 3, our search is sensitive to only
slightly more than half of all potential encounters.  It is much more 
sensitive to brighter (generally more massive) stars for which the search
is essentially complete.  Among the fainter stars for which it is not complete,
it is more sensitive to slower-moving stars.  Up to this point, we have
considered all potential perturbers as equals.  In fact, the stars to which
our search is most sensitive are also potentially the most dangerous.

	To understand why, we first review the Duncan et al.\ (1987) picture
of how a shower is initiated.  During normal times, the action of the planets
depletes comet phase space of all orbits with perihelions $q<a_\nep$, 
where $a_\nep=30\,\au$ is the orbital radius of Neptune.  The most severe
encounter possible, one where this loss cone is suddenly filled, increases
the new-comet rate by a factor $\sim 40$.  (Duncan et al.\ 1987 give this
factor as 20, but we amend their calculation in the appendix
by taking into account more recent work.)\ \  
If the loss cone 
is only partially filled, then the increase in the new-comet rate would be
smaller.  To fill the loss cone requires changing the specific angular 
momentum of the comet by $\Delta J = (2 G M_\odot a_\nep)^{1/2}$.  For
a comet whose aphelion is near the impact parameter $b$ of the passing star,
this requires a change in velocity $\Delta v \gsim \Delta J/b$ (note that
Weissman 1996b finds that shower comets come from the {\it entire} Oort Cloud
rather than primarily those with aphelia close to that of the impact
parameter).  On the
other hand, the velocity change induced by the star on the comet is
$\Delta v = 2 G M_*/(b_cv_*)$ where $M_*$ is the mass of the star, $v_*$ is
its speed relative to the Sun, and $b_c$ is its impact parameter relative
to the comet.  This means that the effective radius of action of the passing
star is
$$b_c \lsim f_* b,\qquad f_* \equiv 
2^{1/2}\,{v_\nep\over v_*}
\,{M_*\over M_\odot},\eqn\bcdef$$
where $v_\nep=5.5\,\kms$ is the orbital speed of Neptune.
Hence, it is only stars that are of order the mass of the Sun and are
moving relatively slowly that can fill most of the loss cone.

	To quantify our sensitivity to the most damaging encounters, we
define a damage parameter, $Q$,
$$Q \equiv (\min\{1,f_*\})^2.\eqn\qdef$$
 From equation \bcdef, it is clear that $Q$ is a rough estimate of the
fraction of the loss cone that is filled.  We estimate the masses of 
main-sequence stars using the empirical mass-luminosity relation of
Henry \& McCarthy (1993).  White dwarfs are assumed to have a mass
$M=0.6\,M_\odot$ and giants to have $M=1\,M_\odot$.

	We evaluate $Q$ for each simulated event in our Monte Carlo 
calculation.  We find for stars passing within $b_\max = 0.05\,\pc$ 
of the Sun within the last $t_d=0.18$ Myr, that only
3\% of events have $Q>0.5$ (of which half have $Q=1$), but we detect 99\%
of these.  An additional 8\% have $0.1<Q<0.5$ of which we detect 98\%.
Another 29\% have $0.005<Q<0.1$ of which we detect 90\%.  It is overwhelmingly
the encounters with very little effect ($Q<0.005$) that escape notice.
These comprise 60\% of all events, only 29\% of which are detected.
If we define a major shower as $Q>0.1$, then we would detect 98\% of 
stars generating such a shower.
Taking account of the fact that we have eliminated 2\% of the stars because
of poor proper-motion measurements, we still would detect 96\%.

	The situation is similar for the less dangerous class of events 
studied in \S~3, passing within $b_\max =0.1\,\pc$ of the Sun during the
last $t_d=0.5$ Myr.
Again, 3\% have $Q>0.5$ of which we detect 98\%, 8\% have $0.1<Q<0.5$ of
which we detect 94\%, and 29\% have $0.005<Q<0.1$ of which we detect 65\%.
The main loss of sensitivity relative to the $b_\max=0.05\,\pc$ case comes
from the 60\% with $Q<0.005$.  For these, our sensitivity drops to 11\%.
	Note that this ``less dangerous'' class of events also includes
all of the ``more dangerous'' ($b<0.05\,\pc$) events occurring up to 0.5 Myr
into the future.

Thus, while it is possible that a close encounter has occurred recently
enough to initiate a comet shower that has not yet arrived, any such
shower will almost certainly be a weak one.  Moreover, it is almost
equally unlikely that a major shower will occur in the next 0.5 Myr.

	This conclusion may seem a bit too reassuring in view of the
fact that in \S~3 we (and Garc\'\i a-S\'anchez et al.\ 1997)
reported a possible {\it future} encounter with
a star which has $V=9.66$ and so is beyond the magnitude limit of
our basic study.  This star has an estimated mass $M\sim 0.64\,M_\odot$
and hence a damage parameter $Q\sim 0.12$.  That is, it is inside the
regime where we claimed to have 98\% sensitivity to {\it past} $b_\max=0.05\,
\pc$ events.  If the sign of the radial velocity had been reversed,
couldn't this star have precipitated a future comet shower but 
(being fainter than the magnitude limit) escaped detection?
The answer is no.
It would have passed the Sun 1.3 Myr ago and so,
if it had precipitated a shower, this shower would have reached the Earth
more than 1 Myr ago.

{\bf Acknowledgements}:   We thank M.\ Duncan, G.\ Newsom, and P. Popowski 
for stimulating discussions and S.\ Tremaine for useful comments. 
The detailed comments by P.\ Weissman on an earlier version 
of this {\it Letter} were particularly helpful.
Work by AG was supported in part by grant AST 94-20746 from the NSF.

\APPENDIX{A}{Normalization of Shower Intensity}

	The dynamical simulations by Duncan et al.\ (1987) determine the 
density profile, $\rho(r)$, of the Oort Cloud, but not its normalization, 
$k$.  Once
the density profile is fixed the maximum absolute intensity of a comet
shower, $I_{\rm shower}$, 
(when a disturbance completely fills the loss cone in the inner Oort
Cloud) is set by the normalization, $I_{\rm shower}\propto k$.
Similarly, for a fixed profile, the steady-state rate 
of new comets, $I_{\rm steady}$, is proportional to the product of the 
normalization and the inverse time scale, $\omega$, 
for comets to diffuse from the Oort Cloud into the loss cone,
$I_{\rm steady} \propto k\omega$.
Hence $I_{\rm shower}/I_{\rm steady} \propto \omega^{-1}$.  This time scale
$\omega^{-1}$ is
estimated theoretically, either analytically or from simulations.
The inverse time scale is the sum of inverse time scales from all relevant
processes.  Duncan et al.\ (1987) argue that two such processes dominate, long
range encounters from stars and Galactic tides, $\omega = \omega_* + \omega_g$.
They show that
$\omega_* = C\rho_*$ and $\omega_g = 0.47 C\rho_0$, where $\rho_*$ is the local
density of stars, $\rho_0$ is the total local density of the disk, and
$C$ is a constant.  Hence,
${I_{\rm shower}/ I_{\rm steady}} \propto (\rho_* + 0.47\rho_0)^{-1}$.
They adopted $\rho_0=0.185\,M_\odot\,\pc^{-3}$ 
(Bahcall 1984) and $\rho_*=0.05\,M_\odot\,\pc^{-3}$.  Recent star counts
show that there are far fewer M dwarfs than previously believed, for
a revised estimate of $\rho_*\sim 0.035\,M_\odot\,\pc^{-3}$.  
Also, Cr\'ez\'e et al.\ (1998)
have recently made a purely {\it local} measurement of the mass density
using accurate distances and proper motions from HIPPARCOS.  They find
$\rho_0 = 0.076\pm 0.015\,M_\odot\,\pc^{-3}$.  If these two newer values are
used
in place of 
Duncan et al.'s\ (1987), the ratio $I_{\rm shower}/I_{\rm steady}$ rises
by a factor $\sim 2$, from 20 to 40.

\endpage

\Ref\bah{Bahcall, J.\ N.\ 1984, ApJ, 276, 169}
\Ref\dunc{Duncan, M., Quinn, T., \& Tremaine, S.\ 1987, AJ, 94, 1330}
\Ref\crb{Casertano, S., Ratnatunga, K.\ U., \& Bahcall, J.\ N.\ 1990, ApJ,
357, 435}
\Ref\crez{Cr\'ez\'e, M., Chereulinst, E., Bienaym\'e, O., \&
Pichoninst, C.\ 1998, A\&A, 329,920} 
\Ref\esa{ESA 1997, The HIPPARCOS Catalogue, ESA SP-1200}
\Ref\gar{Garc\'\i a-S\'anchez, J., Preston, R.\ A., Jones, D.\ L., 
Weissman, P.\ R., Lestrade, J.-F., Latham, D.\ W., \& Stefanik, R.\ P.\ 1997, 
HIPPARCOS Venice '97 Symposium, ESA SP-402, 617}
\Ref\gbf{Gould, A., Bahcall, J.\ N., \& Flynn, C.\ 1996, ApJ, 465, 759}
\Ref\gbft{Gould, A., Bahcall, J.\ N., \& Flynn, C.\ 1997, ApJ, 482, 913}
\Ref\gfb{Gould, A., Flynn, C., \& Bahcall, J.\ N.\ 1998, ApJ, submitted}
\Ref\hmc{Henry, T.\ J.\ \& McCarthy, D.\ W.\ 1993, AJ, 106, 773}
\Ref\hills{Hills, J.\ G.\ 1981, AJ, 86, 1730}
\Ref\hut{Hut, P., Alvarez, A., Elder, W.\ P., Hansen, T., Kauffman, E.\ G., 
Keller, G., Shoemaker, E.\ M., \& Weissman, P.\ R.\ 1987, Nature, 329, 118}
\Ref\ldm{Liebert, J., Dahn, C.\ C., \& Monet, D.\ G.\ 1988, ApJ, 332, 891}
\Ref\mor{M\"ull\"ari, A.\ A., \& Orlov, V.\ V.\ 1996, Earth, Moon, and Planets,
72, 19}
\Ref\sh{Stauffer, J.\ R., \& Hartmann, L.\ W.\  ApJS, 1986, 61, 531}
\Ref\wei{Weissman, P.\ R.\ 1993, BAAS, 25, 1063}
\Ref\weiasp{Weissman, P.\ R.\ 1996a, in Completeing the Inventory of the Solar 
	System, ASP Conference Series, 107, 265}
\Ref\weiemp{Weissman, P.\ R.\ 1996b, Earth, Moon, and Planets, 72, 25}
\Ref\weiny{Weissman, P.\ R.\ 1997, in Near-Earth Objects, ed.\ J.\ L.\ Remo, 
	Annals of the NY Academy of Science, 822, 67}
\Ref\weigar{Weissman, P.\ R., Garc\'\i a-S\'anchez, J., Preston, R.\ A.,
Jones, D.\ L., Lestrade, J.-F., \& Latham, D.\ W.\ 1997, BAAS, 29, 1019}
\Ref\WJK{Wielen, R., Jahreiss, H., \& Kr\"uger, R.\ 1983, IAU Coll.\ 76:
Nearby Stars and the Stellar Luminosity Function, A.\ G.\ D.\ Philip and 
A.\ R.\ Upgren eds., p 163}
\refout
\endpage
\figout
\endpage
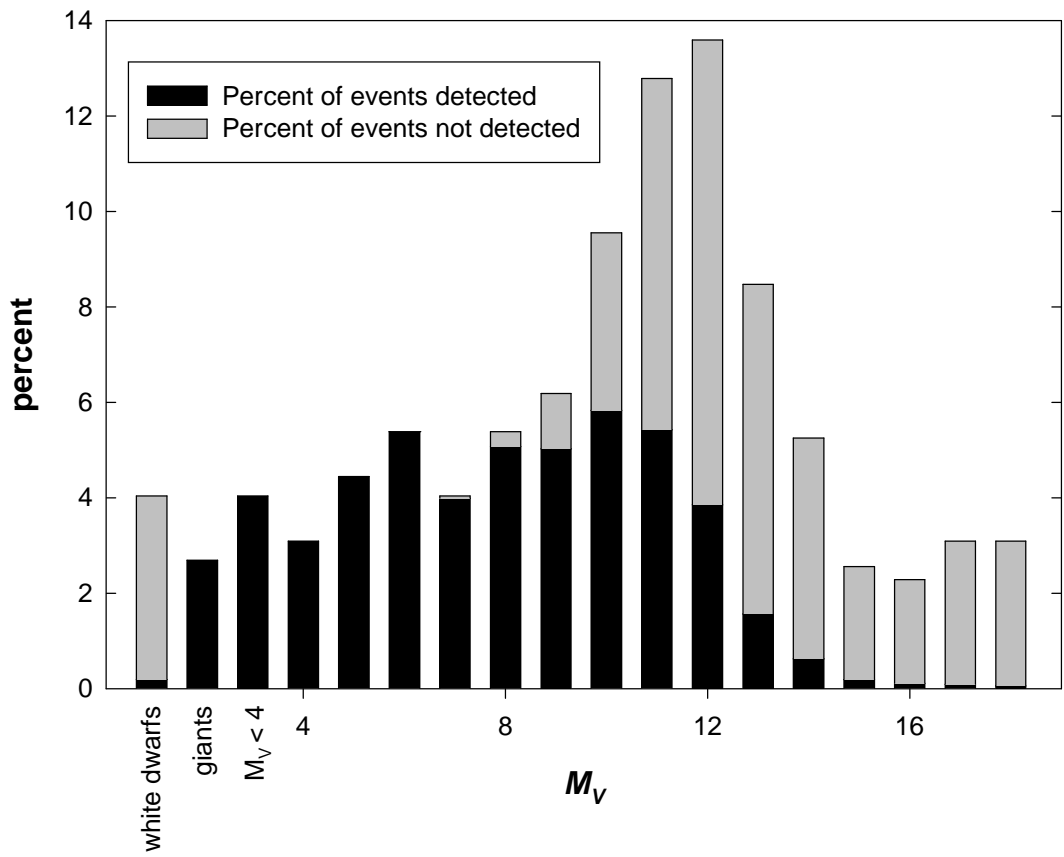
\bye